**The Moisture-Convection Feedback Can Lead to Spontaneous Tropical Cyclogenesis**

Argel Ramírez Reyes* (1) and Da Yang* (1, 2)

Affiliations:

1. University of California, Davis
2. Lawrence Berkeley National Laboratory

*Contact information:

aramirezreyes@ucdavis.edu

dayang@ucdavis.edu

Key points:

1. Spontaneous TC genesis can occur without radiative and surface flux feedbacks.
2. The moisture-convection feedback can lead to spontaneous TC genesis.
3. The MC feedback operates in a time scale of a few hours, which may be efficient enough to help TC genesis from a pre-existing disturbance.

**Abstract**

In contrast to prevailing knowledge, Ramírez Reyes and Yang (2021) showed that tropical cyclones (TCs) can form spontaneously without moisture-radiation and surface-flux feedbacks in a cloud-resolving model (CRM) simulation. Here we ask, why? Thirteen 3D cloud-resolving simulations show that the moisture-convection (MC) feedback can effectively lead to spontaneous TC genesis and intensification in the absence of radiative and surface-flux feedbacks. In the MC feedback, a moister environment favors new deep convective events that further moisten the environment, leading to aggregation of deep convection. The impact of the MC feedback on TC genesis and intensification occurs in two distinct time scales: a short time scale set by detrainment moistening the environment (a few hours) and a long time scale (17 days) due to subsidence drying. The hours-long time scale of detrainment suggests that the MC feedback is an efficient process relevant to TC genesis in the real world.

**Plain language summary**

Computer simulations show that dispersed thunderstorms tend to self-organize into a tropical cyclone (TC) in idealized setups. This process occurs even in highly idealized climates without the most common processes that, in nature, are observed to aid TC formation (e.g., pre-existing disturbance or special spatial structure in ocean energy fluxes and radiative energy fluxes). What causes tropical cyclones to form in such an exotic atmosphere? We find, using computer simulations, that thunderstorms and environmental moisture reinforce one another, causing disorganized patches of storms to aggregate into a tropical cyclone. We call this the moisture-convection (MC) feedback. By varying our simulation parameters, we find that the MC feedback operates over periods of a few hours to several weeks. The MC feedback's fast (hours-long) component may be important for understanding and forecasting the genesis of real-world TCs.

## 1. Introduction

In Earth's atmosphere, there are around 90 tropical cyclones (TCs) every year. They are warm-cored, rapidly rotating storms with horizontal scales of the order of 500 km in maturity and are typically formed over tropical oceans. Despite extensive research and improved forecast techniques, uncertainties in forecasting TC genesis and intensification hinder safety and evacuation planning by policymakers. TC-associated intense rainfall, strong winds, and storm surges continue to cost lives and cause economic losses and other societal impacts. Therefore, improving our understanding of TC genesis and intensification is of primary importance in tropical meteorology and atmospheric science.

In computer simulations, TCs can self-emerge from random convection over uniform sea-surface temperatures on an *f*-plane (Bretherton et al., 2005; Carstens & Wing, 2020; Davis, 2015). This phenomenon is known as *spontaneous TC genesis*. On Earth, TCs often develop from a pre-existing disturbance, and a vast amount of literature has explored the leading-order processes that dominate typical TC genesis events (Dunkerton et al., 2009; Emanuel, 1986, 2018; Montgomery et al., 2006; Ooyama, 1982; Raymond et al., 2007). However, the processes that lead to spontaneous TC genesis in these idealized setups are still present in the real atmosphere and may help the evolution of a pre-existing disturbance into a TC. Therefore, spontaneous TC genesis offers a simplified framework to understand some elements of TC genesis (in particular the up-scale organization of convection) without the full complexity of the real atmosphere (Davis, 2015). Wing et al. (2016) showed that spontaneous TC genesis requires either radiative feedback or surface-flux feedback. This result was then confirmed and recast by Muller and Romps (2018) using small-domain (1000 km x 1000 km) cloud-resolving simulations. However, Ramírez Reyes and Yang (2021) showed that TCs can self-emerge without radiative and surface flux feedbacks using the same cloud-resolving model (CRM). A major difference is that Ramírez Reyes and Yang (2021) used a higher Coriolis parameter, making TCs smaller and easier to fit into a limited computing domain. This result challenges the prevailing understanding of spontaneous TC genesis.

What leads to spontaneous TC genesis and intensification without radiative and surface flux feedbacks? Here, we test the hypothesis that the moisture-convection (MC) feedback leads to TC genesis and intensification in the study of Ramírez Reyes and Yang (2021). The moisture convection feedback requires horizontal moisture gradients which influence and are in turn influenced by the spatial distribution of deep convection. In the MC feedback, deep convection moistens its environment. The moister environment promotes new convective events in the same region, which further moisten the environment, promoting the next cycle of convection. The mutual reinforcement of convection and moisture could help TC genesis in more than one way. For example, promoting the emergence of deep convective towers can amplify ambient vorticity by tilting and stretching (Montgomery et al., 2006) and can create a moistening tendency in the lower troposphere, which weakens the evaporation of rain and its associated divergence in the low levels (Nolan, 2007; Raymond et al., 2007; Wang, 2012). The MC feedback was first proposed by Scorer and Ludlam (1953), and it recently gained relevance from the thorough studies of Tompkins (2001; Tompkins & Semie, 2017). Since then, several authors have studied the impact of a MC feedback in organizing convection both in the small scales (~5-50 km) and in the larger scales (~500 km and above) (G. C. Craig & Mack, 2013; Grabowski & Moncrieff, 2004; Kuang & Bretherton, 2006; Seeley & Romps, 2015; Tompkins & Semie, 2017; Waite & Khouider, 2010; Wang, 2014b; Yang, 2019).

In this work, we use a mechanism-denial experiment and high-resolution simulations to test the hypothesis that the MC feedback is responsible for spontaneous TC genesis and intensification without radiative and surface-flux feedbacks in the simulations of Ramírez Reyes and Yang (2021). We also explore the time scales in which the components of this feedback operate, and we discuss if the MC feedback could be relevant to real TC genesis.

## 3. Hypothesis: the MC feedback leads to spontaneous TC genesis in the absence of radiative and surface-flux feedbacks

2.1 The Moisture-convection feedback

Dry environments inhibit deep convection, and moist environments favor deep convection. Reciprocally, deep convection moistens its environment, which favors new deep convective events, closing a feedback loop. This simple but intuitive statement summarizes the MC feedback on the small scales (~5-50 km) and large scales (~500 km and above). The difference between the two cases resides in which processes dominate the transport of moisture in the free troposphere and what are their associated time scales. For example, close to the convective cloud, the mixing of cloudy air into the environment occurs on timescales associated with "fast" mixing processes, while in the larger scale moisture is transported via "slow" advective processes (Grabowski &

Moncrieff, 2004). Because we are trying to understand the up-scale organization of convection into a TC, both spatial scales and their associated time scales may be relevant for spontaneous TC genesis: the small-scale processes may be relevant during the genesis stage, and the large-scale processes will become relevant once the large-scale circulations of a TC are established. After giving an overview of the MC feedback, in the following paragraphs we will try to establish two relevant time scales (and their associated spatial scales) in which the interaction between moisture and convection may be relevant to spontaneous TC genesis.

A deep convective cloud that rises with vertical velocity $W_c$ exchanges mass with its environment through mixing processes, by which environmental air flows into the cloud and cloudy air flows into the environment. These mixing processes have diverse spatial scales, some are small scale eddies and others occur through larger scale flows. In the following, we will refer to these broadly defined mixing processes as entrainment and detrainment, acknowledging that their usage for an evolving convective plume may be subject to discussion (de Rooy et al., 2013; Yano, 2014). These mixing processes are characterized by fractional detrainment and entrainment rates of $\delta$ and $\epsilon$, respectively. Mixing environmental air into convective updrafts reduces the buoyancy of the convecting parcels due to evaporative cooling (Tompkins & Semie, 2017) and the buoyancy effect of water vapor (Seidel & Yang, 2020; Yang & Seidel, 2020). This implies that mixing moist environmental air into the updrafts results in a smaller buoyancy reduction than it would if the environmental air were drier. Through similar mixing processes (the turbulent eddies adding moisture to the immediate environment of the clouds and the mixing of the moisture of multiple convective plumes into the large scale by coherent large-scale flows), moist convection influences moisture in the environment. Moisture detrained from convective clouds into the environment creates an anomalously moist region that favors new deep convective events in the vicinity of existing convective clouds and around the location of recently extinct convective clouds. Outside of the deep convective clouds, where moisture is not added directly by detrainment and entrainment but is advected by large-scale flows, other slower processes affect moisture: Inthe clear-sky environment, air subsides with a characteristic vertical velocity $W_s$. Subsidence dries the atmosphere, inhibiting new deep convective events.

This reciprocal interaction of environmental moisture and convection constitutes a feedback process that promotes the organization of deep convection on the scale of a few convective storms and on large scales. To the best of our knowledge, this feedback was first proposed by Scorer and Ludlam (1953) and discussed by Randall and Huffman (1980). It has been thoroughly examined by Tompkins (2001; Tompkins & Semie, 2017) in studies of convective self-aggregation; this feedback was then further developed and applied to study the convective organization of the tropical atmosphere in the scale of individual convective clouds (Waite & Khouider, 2010), in the large-scale (Grabowski & Moncrieff, 2004) and in the scales in-between (G. C. Craig & Mack, 2013; Kuang & Bretherton, 2006; Seeley & Romps, 2015; Yang, 2019) and by Wang (2014b) in the study of TC genesis in numerical models. Below we analyze two characteristic timescales associated with this feedback; both timescales contribute to generating horizontal moisture gradients in different scales.

2.2 Entrainment and detrainment timescale

The cloudy air ascending at vertical velocity $W_c$ will exchange mass with the environment with a fractional detrainment rate δ (with units of 1/m). This detrainment rate corresponds to a vertical length $\text{Length}_\delta \equiv \frac{1}{\delta}$ over which convection moistens the environment. A typical value of the fractional detrainment rate of deep convection on Earth is 0.2 km$^{-1}$ (Romps, 2014). We select a characteristic vertical velocity of the convecting plume of 1 ms$^{-1}$, which is approximately the mean updraft velocity from the Control simulation described below. This value is also close to the median vertical velocity of updrafts in some observations of tropical deep convection (Lucas et al., 1994). Another proposal would be to take the mean vertical velocity in the simulation (e.g., Fig 3 in (Wang, 2014a)). However, this value would be zero in RCE. Therefore, we consider that using only the upward motion represents well the role of convective updrafts in the moisture exchange. Using our selected vertical velocity, we find that the moistening time scale is

$$\tau_d = \frac{\text{Length}_\delta}{W_c} \approx \frac{5000 \text{ m}}{1 \text{ ms}^{-1}} \approx 1.38 \text{ hours.}$$

Here we assume that, in the leading order, the entrainment and detrainment rates are approximately equal (Romps, 2014). We expect this time scale to be more relevant in the small scale.

2.3 Subsidence timescale

In clear-sky regions away from organized convection, air subsides with a characteristic velocity $W_s$. This process dries the clear-sky region and tends to promote the convective organization in longer temporal scales than the previously described detrainment processes. For our experiments, it is sensible to select the radiative subsidence associated with the outer region of a TC. In the outer region of a TC, subsidence has characteristic velocities of around 2 mm s$^{-1}$ (Chavas et al., 2015). A characteristic length of the problem is the scale height of water vapor which has values around 3000 m (Romps, 2014). With these two ingredients, we can estimate the time scale of drying by subsidence:

$$\tau_s = \frac{\text{Water vapor scale height}}{W_S} \approx \frac{3000 \text{ m}}{0.002 \text{ m s}^{-1}} \approx 17 \text{ days.}$$

We note that $\tau_s$ is much longer than $\tau_d$, suggesting that detrainment of water vapor is a more efficient process in creating horizontal moisture perturbations. Because we are using a mean subsidence rate for a region outside of a TC, we expect this timescale to be relevant for the large-scale processes (once a TC with its associated large-scale circulations is already formed).

2.4 Expectations

We hypothesize that the MC feedback is responsible for spontaneous TC genesis in the absence of a pre-existing disturbance or other feedbacks. The mutual reinforcement of moisture and convection makes the emergence of deep convective storms more likely. Deep convection in a vorticity-rich environment would produce local vorticity anomalies by vortex tilting and stretching. These anomalies are then candidates for vortex merger (Montgomery et al., 2006). The same moistening tendency would also help create a thermodynamic profile that inhibits evaporation and its associated divergence in the lower levels of the troposphere (Raymond et al., 2007). Our hypothesis predicts that the homogenization of clear-sky water vapor on a *shorter* time scale than $\tau_d$ will effectively remove horizontal water vapor gradients. This removes a factor that promotes the emergence of nearby deep convective events and thereby inhibits spontaneous TC genesis (P1).

Additionally, subsidence of dry air inhibits convection outside the core of a TC with a time scale of $\tau_s$. Inhibiting convection outside the core allows convection to be concentrated near the core, favoring the intensification of TCs. Therefore, we expect that moistening the descending branch on a time scale faster than $\tau_s$ will prevent the intensification of the spontaneously formed TCs (P2).

## 3. Methods

3.1 The cloud-permitting model and simulation details

We run 3D simulations using the *System for atmospheric modeling* (SAM, version 6.10.10) (Khairoutdinov & Randall, 2003). SAM solves the anelastic equations of motion using a finite-difference numerical scheme in an Arakawa-C grid. Regarding thermodynamics, SAM solves the conservation equation for frozen moist static energy. We use the CAM radiation scheme for long-wave and short-wave fluxes (Collins et al., 2004), and the default SAM single-moment bulk microphysics scheme. The latent and sensible heat fluxes are computed using a bulk formulation where the transfer coefficients are computed using the Monin-Obukhov theory with code adapted from the Community Climate Model 3 (Kiehl et al., 1996). We used the SAM Smagorinsky SGS parameterization described in (Deardorff, 1980). One difference between SAM and other models like WRF (Skamarock et al., 2021) is that SAM does not employ a specific parameterization for the boundary layer. Instead, the sub grid-scale fluxes in the boundary layer are computed by the same SGS scheme. To prevent erroneous values near the model surface, only the vertical grid scale is used on the SGS grid length when the horizontal grid spacing is much larger than the vertical (Khairoutdinov & Randall, 2003). We output 3D fields every two hours and 2D fields every hour.

The simulation domain is a square of 1024 km x 1024 km in the horizontal dimensions and 34.8 km in the vertical. Grid spacing is 2 km in the horizontal direction and stretched in the vertical direction: it is 50 m from $z = 0$ m to $z = 1050$ m. Then, the vertical grid spacing increases gradually until z = 3000 m, where it becomes 600 m. The boundary conditions are periodic in the horizontal directions. The bottom boundary is an ocean with a fixed surface temperature of 300 K, and the upper 30 levels are a sponge layer to prevent the reflection of gravity waves. We use a constant Coriolis parameter of $f = 5 \text{x} 10^{-4} \text{ s}^{-1}$. This value is ten times that of 20° latitude on Earth. A similar value has been used before (Cronin & Chavas, 2019; Khairoutdinov & Emanuel, 2013; Ramírez Reyes & Yang, 2021). The increased Coriolis parameter is necessary for spontaneous TC genesis without radiative and surface flux feedbacks in this domain with a grid spacing of 2 km (Ramírez Reyes & Yang, 2021), and it also allows for the reduction of the size of individual TCs, allowing several TCs to exist in this domain (Khairoutdinov & Emanuel, 2013). While in real Earth, radiative and surface-flux feedbacks may be necessary for TC genesis to occur, turning them off allows us to explore in detail the role of the MC feedback. The knowledge gained from this framework can then be compared to more realistic simulations. We initialize the simulations using a sounding from the last days of a non-rotating RCE simulation, the same profile used in (Yang, 2018). In all the simulations, we disable the radiative and surface-flux feedbacks by substituting surface heat fluxes and radiative fluxes with their horizontally averaged value before applying them (Muller & Romps, 2018; Ramírez Reyes & Yang, 2021; Wing et al., 2016). By removing the spatial variation of surface fluxes, we prevent surface fluxes from creating and intensifying moisture gradients, which effectively disables the surface-flux feedbacks, especially the so-called WISHE mechanism (Montgomery & Smith, 2014).

3.2 Experiment design: Weakening the MC feedback

We design experiments to test P1 and P2 by homogenizing clear-sky water vapor. We run in total 13 simulations. Our Control simulation is similar to the HomoAll experiment of Ramírez Reyes and Yang (2021), where radiative and surface-flux feedbacks are disabled by substituting radiative heating rate and surface fluxes with their horizontal averages before applying them (Muller & Romps, 2018; Ramírez Reyes & Yang, 2021; Wing et al., 2016). Although we expect the buoyancy effect of water vapor on buoyancy of the convecting cloud to be small when compared to the effects of energy released by phase changes (Yang, 2018), we turn off the buoyancy effect of water vapor as in (Yang, 2019) to further simplify the system. We then setup our mechanism-denial experiment.

This experiment consists of a set of simulations that differ from Control by relaxing (nudging) the clear-sky (unsaturated grid points) water vapor toward its horizontal average. In the anomalously moist regions, homogenizing clear-sky water vapor means drying; in the anomalously dry regions, it is moistening. To homogenize clear-sky water vapor, we add a relaxation term to the water vapor field in the unsaturated grid points as done by Yang (2019).

$$\widetilde{\delta r}(i,j,k,t_i)\Delta t = \begin{cases} \delta r(i,j,k,t_i)\,\Delta t & \text{if } r_{cond}(i,j,k,t_i) > 0.01 \text{ g kg}^{-1} \\ \delta r(i,j,k,t_i)\,\Delta t + \big(\bar{r}(k,t_i) - r(i,j,k,t_i)\big)\Delta t/\tau_r & \text{otherwise} \end{cases},$$

where $i, j,$ and $k$ represent the grid indices in the space dimensions and $t_i$ represents the time index. $\delta r$ is the original tendency of the water vapor field computed by SAM and multiplied by the timestep $\Delta t$. $r_{cond}$ is the total condensate mixing ratio. The overbar represents an average over all the points that fulfill the same criterion of low condensate. $\tau_r$ is the relaxation time scale. We let $\tau_r$ take 12 values: 0.5 h, 1h, 3h, 5h, 8h, 12h, 1 day, 2 days, 5 days, 8 days, 10 days and 15 days. This method resembles the relaxation done by Seeley and Romps (2015), except that they nudged the relative humidity field. Like the method employed by Seeley and Romps, our method effectively reduces moisture gradients (Figure 1).

3.3 Detection and characterization of TC-associated inflow and updrafts

To characterize the location of convective updrafts relative to the TC centers and the intensity of the inflow we first identify TC centers as local minima of surface pressure at each time step during each 10-day period of each simulation. Next, we compute the total number of updrafts (defined as grid points where the vertical velocity exceeds 2 ms$^{-1}$,) that occur at each distance from the center (using 2-km bins starting in r = 1 km) and divide it by the number of points that fall within said bin (a measure of the area of the region that falls between said bins). Finally, we divide the total number of updrafts that occur between $r = 3r_{max}$ and r = $6r_{max}$ by those that appear between $r = 0$ and $r = 6r_{max}$. This quantity, which we call the "updraft ratio" shows how concentrated convection is near the radius of maximum winds. We selected the region between $r = 3r_{max}$ and r = $6r_{max}$ somewhat arbitrarily. However, we tested other choices of radii, and the results remain robust. TCs that have most of their convective events near the radius of maximum winds would have an updraft ratio close to zero, and TCs that have as many updrafts close to the radius of maximum winds as far from it would have an updraft ratio close to 0.5. To compute the inflow we create an "average" TC by aligning each center and averaging the velocity fields, as in (Ramírez Reyes & Yang, 2021). We then compute the mean radial velocity in a region bounded by the radius of maximum winds ($r_{max}$) and $6r_{max}$ in the radial direction and from the surface to 6 km height. We perform the calculation for all the experiments with TCs for 10-day periods between day 60 and 100 (e.g., day 60 to 70, 70 to 80, 80 to 90 and 90 to 100).

4. Results

Relaxation of clear-sky water vapor to its horizontal average has little effect on the domain-mean specific humidity ( ~16g/kg in Control vs 14g/kg in the experiment with $\tau_r = 0.5$ h) but a larger effect in the standard deviation of specific humidity (2.1 g/kg in Control vs 1.2 g/kg kg in the experiment with $\tau_r = 0.5$ h ) (Figure 1). This smaller standard in the experiments with nudging implies a more homogeneous water vapor field. Furthermore, as the nudging becomes faster, the standard deviation becomes smaller.

Homogenizing clear-sky water vapor on a timescale shorter than $\tau_d$ prevents spontaneous TC genesis (P1). TCs spontaneously emerge in Control and in simulations with nudging timescale $\tau_r$ between 2 hours and 15 days. However, no TCs form in simulations with nudging timescales smaller than 2 hours. In Figure 2 we show snapshots of precipitable water (PW), surface wind speed, and surface pressure at day 95 for a selected subset of simulations (see also Movie S1). We observe that the PW shows negligible spatial variance for the short relaxation time scales (e.g., $\tau_r = 0.5$ hours or $\tau_r = 1$ hour), which increases as $\tau_r$ becomes longer. In Control and $\tau_r = 15$ days, there are distinct large-scale high PW regions that coincide with the TCs, which is surrounded by low PW regions. For relaxation time scales below 2 hours, we observe randomly distributed winds with maximum speeds of around 10 m s$^{-1}$. When $\tau_r$ increases above 2 hours, we observe regions of increased surface wind speed around the eye. TCs in the experiments with $\tau_r$ between 3 hours and 5 days show modest intensities by day 60, while experiments with relaxation time scales above 10 days show similar intensities to TCs in Control.

Our experiments also support P2 and show that homogenizing clear-sky water vapor on a multi-day time scale prevents the intensification of spontaneously formed TCs. The behavior of simulations can be described as three distinct regimes: No TCs, weak TCs, and intense TCs (similar in intensity to those in Control). Figure 3 shows the maximum surface wind speed and minimum surface pressure in the domain versus time for a selected subset of experiments (see Figure S2 for all the experiments). The experiments with $\tau_r$ below 2 hours show almost constant maximum wind speeds of around 15 ms$^{-1}$ (corresponding to random wind gusts) and nearly constant minimum surface pressures of around 1000 hPa. These simulations show random convection. Experiments with $\tau_r$ higher than 2 hours show stronger wind speeds and lower pressures at the surface: after an initial intensification period, simulations with $\tau_r$ between 3 hours and 5 days show surface wind speeds and surface pressures fields that oscillate around 20 ms$^{-1}$ and 990 hPa, corresponding to weak TCs. Finally, the experiment with $\tau_r$ equal to 15 days mimics the Control simulation, with maximum wind speeds oscillating around 40 ms$^{-1}$ and minimum surface pressures that oscillate around 960 hPa after an intensification period. These three regimes ("short" relaxation time scales with no organization, "moderate" relaxation time scales with weak TCs and "long" relaxation time scales with strong TCs) suggest that the coupling between environmental water vapor and convection may operate in two different mechanisms in TC genesis and TC intensification. A fast process is instrumental for initial convective organization and spontaneous TC genesis, and a slower process helps TC intensification. These results are consistent with P2. We note though, that the simulations with $\tau_r = 5$ days and $\tau_r = 10$ days reach similar intensities to Control by the end of the 100-day period (Figure S1). This suggests that the transition between the no non-intensifying and the intensifying regime may not be abrupt. A more detailed study of the transition would require a substantial increase in computational expense and is left to future work.

Here we speculate about how clear-sky water vapor homogenization on a multi-day time scale prevents the intensification of TCs. Intensification of TCs requires a strong inflow to bring air parcels closer to the eyewall while nearly maintaining their angular momentum. This amplifies the tangential wind speeds and intensifies the TC (Montgomery & Smith, 2014). Meanwhile, drying by subsidence beyond the eyewall opposes the appearance of convection beyond the TC eyewall. In our experiments, homogenizing water vapor leads to the moistening of the subsiding areas, allowing deep convective events beyond the radius of the eyewall. This increased convection beyond the eyewall reduces the inflow to the eyewall and therefore opposes the intensification of the TCs.

To test this hypothesis, Figure 4 shows the relationship between mean inflow and updraft ratio, which measures how concentrated updrafts are near the radius of maximum tangential winds (a value close to 0.0 shows that convection is concentrated near the radius of maximum winds and a value of 0.5 shows that convection is uniformly distributed). In our experiments, the updraft ratio is related to the mean inflow approximately by $log(\text{inflow}) = -2.77 * \text{updraft ratio}$ (correlation coefficient = -0.77), showing that TCs with most updrafts concentrated near $r_{max}$ have a stronger inflow than those where convection is uniformly distributed at all radii, which is consistent with our hypothesis. However, we note that this is a diagnostic result and does not indicate causality. For example, it does not prove that the reduced inflow is caused by the uneven distribution of convection. Furthermore, another study found that moistening of the subsidence region of TCs with a long time scale results in a larger steady-state intensity than in a simulation with no moistening in an axisymmetric model (Rousseau-Rizzi et al., 2021). Thus, a reduction in the moisture at large radii would have two competing effects: it promotes convective organization by suppressing convection outside of the core, but it also reduces the moisture supply to the storm, which is important for development (Fritz & Wang, 2014). Other plausible explanations include that asymmetries in convection outside the eyewall may lead to the weakening of the TCs (Nolan et al., 2007) or that the environmental maximum potential intensity of TCs is reduced by modification of the moisture field (Bister & Emanuel, 1998). However, computation of the potential intensity with the routine provided by Bister and Emanuel (Bister & Emanuel, 2002) does not support this hypothesis (see Figure S2 in the supplemental material). A conclusive investigation of the causes for the different intensities when weakening the MC feedback on the "long" timescales would require a deeper study.

## 5. Main findings and implications

This work answers what process is responsible for spontaneous TC genesis in the absence of radiative and surface-flux feedbacks in the simulations of Ramírez Reyes and Yang (2021). We perform mechanism-denial experiments using cloud-permitting simulations to test the hypothesis that the MC feedback is responsible for spontaneous TC genesis and intensification in the absence of radiative and surface-flux feedbacks. In the MC feedback, moisture detrained from convective clouds moistens the environment and makes it favorable for new convective clouds, and drying by subsidence inhibits convection in the region. Both processes can lead to the convective organization (Tompkins, 2001). We find that the moisture gradients created by the MC feedback lead to spontaneous TC genesis and intensification in the absence of radiative and surface-flux feedbacks.

The MC feedback influences spontaneous TC genesis and intensification in two different time scales. Using dimensional analysis, we estimate that the entrainment and detrainment time scale is about a few hours, setting the short time scale of the MC feedback. We test this prediction by weakening the MC feedback across different time scales. We observe that using a time scale shorter than 2 hours, spontaneous TC genesis is inhibited, which corroborates our hypothesis. This "fast" time scale should not be mistaken to indicate the time to genesis. Instead, this fast time scale suggests that the MC feedback is efficient and relevant for real-world TC genesis as an aide to the process of converting a pre-existing disturbance into a Tropical Cyclone. For example, this result seems consistent with the real case study of Wang (2014b), who pointed out that repeated congestus convection moistens the lower troposphere, leading to organized convection and to TC genesis. This result and previous literature highlight the importance of continuing to evaluate and improve the representation of moisture and its interaction with convection, especially in general circulation models (GCMs), which remain a primary tool for TC genesis forecasting and do not explicitly resolve convection (Halperin et al., 2020).

Subsidence drying sets a time scale of ~ 17 days. We propose that subsidence drying promotes intensification of the TC by preventing the appearance of new convective events far from the radius of maximum winds, which would reduce the radial advection of angular momentum required for intensification of the TC (Montgomery & Smith, 2014). When we homogenize clear-sky water vapor on a multi-day time scale, we observe more frequent updrafts far from the eyewall of incipient TCs. This is accompanied by reduced inflow and lower TC intensity when compared with experiments with longer relaxation time scales or with no relaxation. This result again is consistent with our hypothesis. However, there are other possible explanations for the decreased TC intensity when using shorter water vapor-nudging timescale. For example, vertical wind shear (Alland et al., 2017), advection of dry environmental air (Wu et al., 2015), and moisture-radiation feedbacks (Ruppert et al., 2020) can all affect TC intensity. Examining these mechanisms in our modeling framework will require future research.

**Data and software availability:**

Following the Best Practices for Preservation and Replicability (Schuster et al., 2022), we make available the code and instructions to replicate our results. the system for atmospheric modeling (SAM) is free software and can be obtained from http://rossby.msrc.sunysb.edu/~marat/SAM.html . The version used in this work together with modified source code, initialization profiles, namelists and data analysis scripts are hosted on Zenodo and can be found with the digital object identifier (DOI) 10.5281/zenodo.6978839 (Ramírez Reyes & Yang, 2022). The data was analyzed using the Julia Programming Language (Bezanson et al., 2017) available under MIT license at www.julialang.org, with the Images.jl package available with MIT license at https://github.com/JuliaImages/Images.jl. Figures were created using the Makie.jl package (Danisch & Krumbiegel, 2021) available under MIT license at https://github.com/JuliaPlots/Makie.jl.

**Appendix 1**

On the importance of the SGS parameterization at a grid spacing of 2 km

This experiment consists of one run that differs from Control only by turning off the horizontal mixing of water vapor fields from the SGS parameterization. This simulation will be referred to as NoSGS. In this study, we use a grid spacing of 2 km. At this resolution, we would expect a significant part of the entrainment and detrainment from clouds still depends on the SGS parameterization (Bryan et al., 2003; George C. Craig & Dörnbrack, 2008; Tompkins & Semie, 2017). By turning off the SGS parameterization, we remove the unresolved portion of eddy transports, specifically the entrainment and detrainment, allowing us to directly evaluate the hypothesis that a reduction of detrainment and entrainment of clouds would inhibit TC genesis (also consistent with (Zhao et al., 2012)). However, we discovered that spontaneous TC genesis still occurs in this setup. When taking a close look at the moisture fields, Figures 1 and 2 show that turning off the horizontal mixing of water vapor by the SGS parameterization does not have a strong effect on the standard deviation of water vapor. This could imply that SAM has substantial horizontal mixing at a grid spacing of 2 km, rendering the contribution of the SGS parameterization less important. The mixing could come from resolved flows or the numerical implementation of SAM. SAM uses the finite difference method, which might be diffusive and act as an effective way of mixing water vapor. All the cited works used different numerical models. Therefore, our finding does not directly oppose theirs. We consider that the representation of mixing processes in CRMs and how these processes depend on the resolution is a problem worth revisiting in future studies.

**Acknowledgments**

This work is supported by a Packard Fellowship for Science and Engineering, NSF CAREER Award, and the France-Berkeley Fund to Da Yang. Argel Ramírez Reyes was also supported by a CONACYT-UCMexus fellowship. Argel Ramírez Reyes thanks Matthew R. Igel, and Adrian Tompkins and Chis Davis for their useful input in this work.

**Submitted for consideration at the Geophysical Research Letters**Kiehl, T., Hack, J., Bonan, B., Boville, A., Briegleb, P., Williamson, L., & Rasch, J. (1996). Description of the NCAR Community Climate Model (CCM3). https://doi.org/10.5065/D6FF3Q99

Kuang, Z., & Bretherton, C. S. (2006). A Mass-Flux Scheme View of a High-Resolution Simulation of a Transition from Shallow to Deep Cumulus Convection. *Journal of the Atmospheric Sciences*, *63*(7), 1895–1909. https://doi.org/10.1175/JAS3723.1

Lucas, C., Zipser, E. J., & Lemone, M. A. (1994). Vertical Velocity in Oceanic Convection off Tropical Australia. *Journal of the Atmospheric Sciences*, *51*(21), 3183–3193. https://doi.org/10.1175/1520-0469(1994)051<3183:VVIOCO>2.0.CO;2

Montgomery, M. T., & Smith, R. (2014). Paradigms for tropical cyclone intensification. *Australian Meteorological and Oceanographic Journal*, *64*(1), 37–66. https://doi.org/10.22499/2.6401.005

Montgomery, M. T., Nicholls, M. E., Cram, T. A., & Saunders, A. B. (2006). A Vortical Hot Tower Route to Tropical Cyclogenesis. *Journal of the Atmospheric Sciences*, *63*(1), 355–386. https://doi.org/10.1175/JAS3604.1

Muller, C. J., & Romps, D. M. (2018). Acceleration of tropical cyclogenesis by self-aggregation feedbacks. *Proceedings of the National Academy of Sciences*, *115*(12), 2930–2935. https://doi.org/10.1073/pnas.1719967115

Nolan, D. S. (2007). What is the trigger for tropical cyclogenesis? *Australian Meteorological Magazine*, *56*(4), 241–266.

Nolan, D. S., Moon, Y., & Stern, D. P. (2007). Tropical Cyclone Intensification from Asymmetric Convection: Energetics and Efficiency. *Journal of the Atmospheric Sciences*, *64*(10), 3377–3405. https://doi.org/10.1175/JAS3988.1

Submitted for consideration at the Geophysical Research Letters

1.

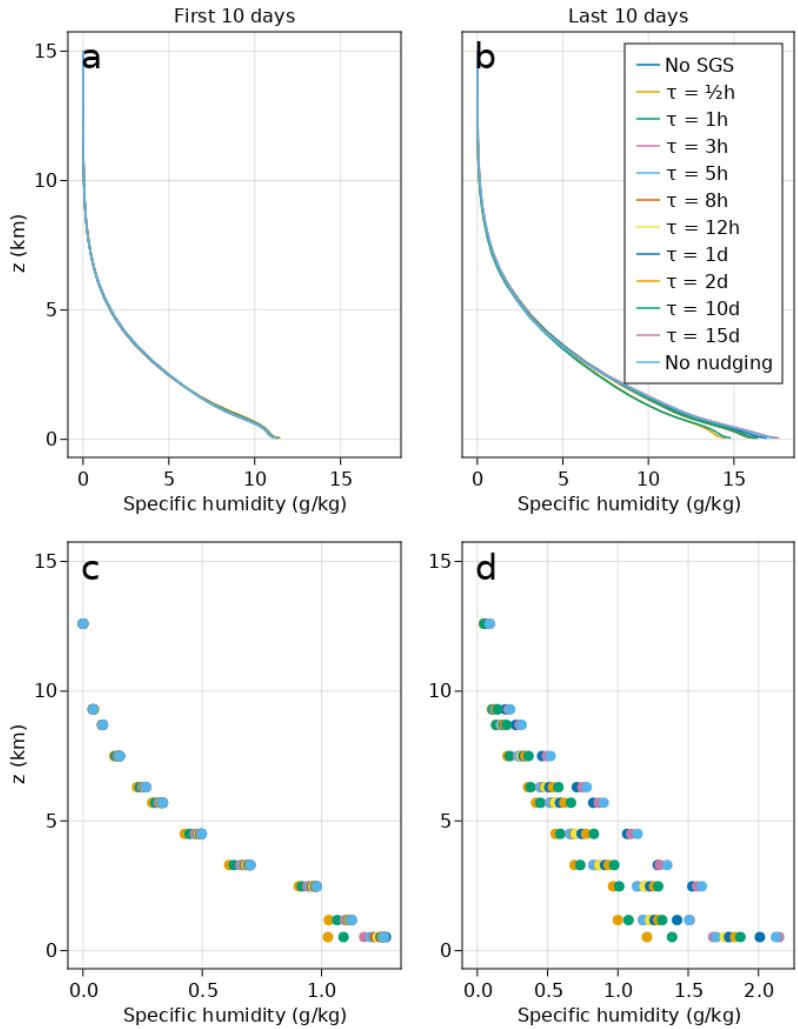

Figure 1. Mean profile of specific humidity averaged over the first 10 days of the simulation (a) and the last then days of the simulation (b). The standard deviation of specific humidity averaged over the first 10 days of the simulation (c) and the last ten days of the simulation (d). See Appendix 1 for a note on the experiment NoSGS.

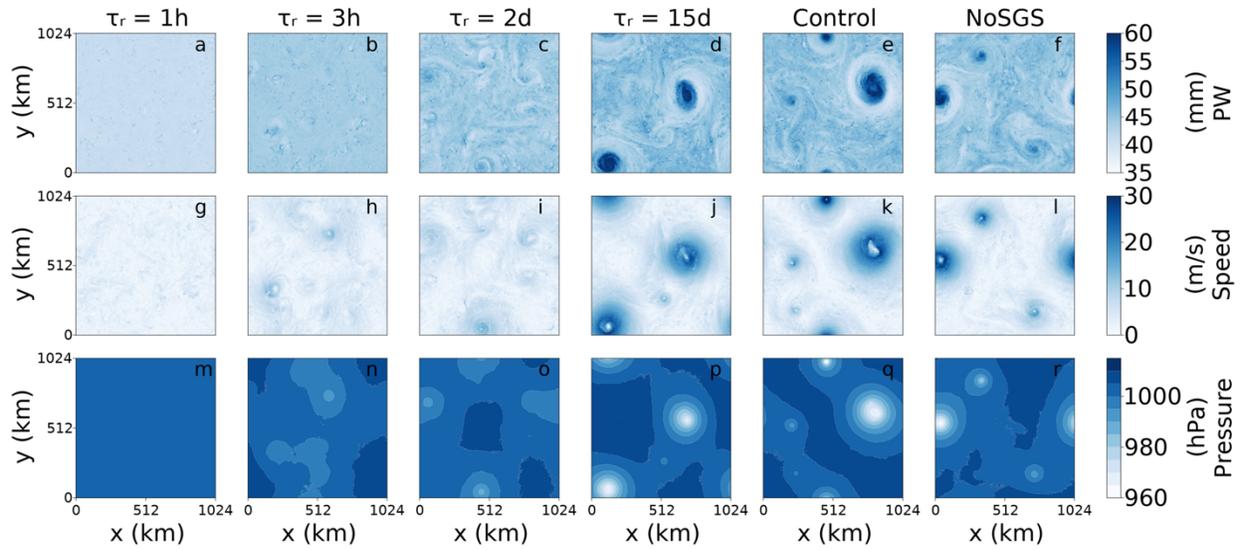

Figure 2. Map views of precipitable water, surface pressure, and surface wind speed at $t = 95$ days (snapshot). (a-f) precipitable water (mm), (g-l) surface wind speed (m s$^{-1}$), and (m-r) surface pressure (hPa). Columns 1-4 correspond to experiments with water vapor relaxation time scales of 1 h, 3 h, 2 days, and 15 days respectively, and the fifth column is the Control experiment and the sixth column is the NoSGS experiment (see Appendix 1). Simulations with 2-km grid spacing.

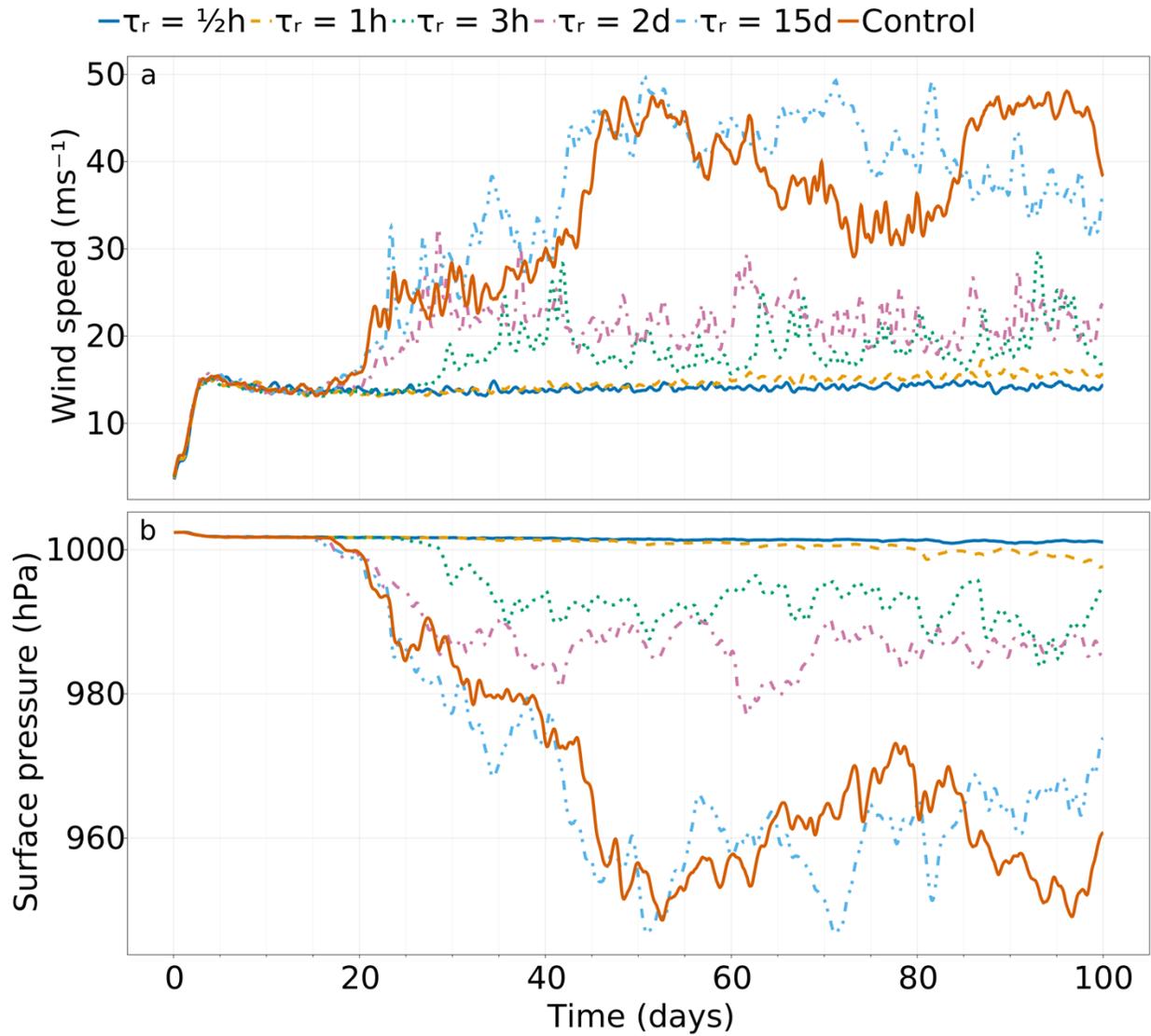

Figure 3. Time evolution of the maximum surface wind speed (m/s) (a) and minimum surface (hPa) (b). Hourly data is smoothed with a moving average filter with window = 12 hours.

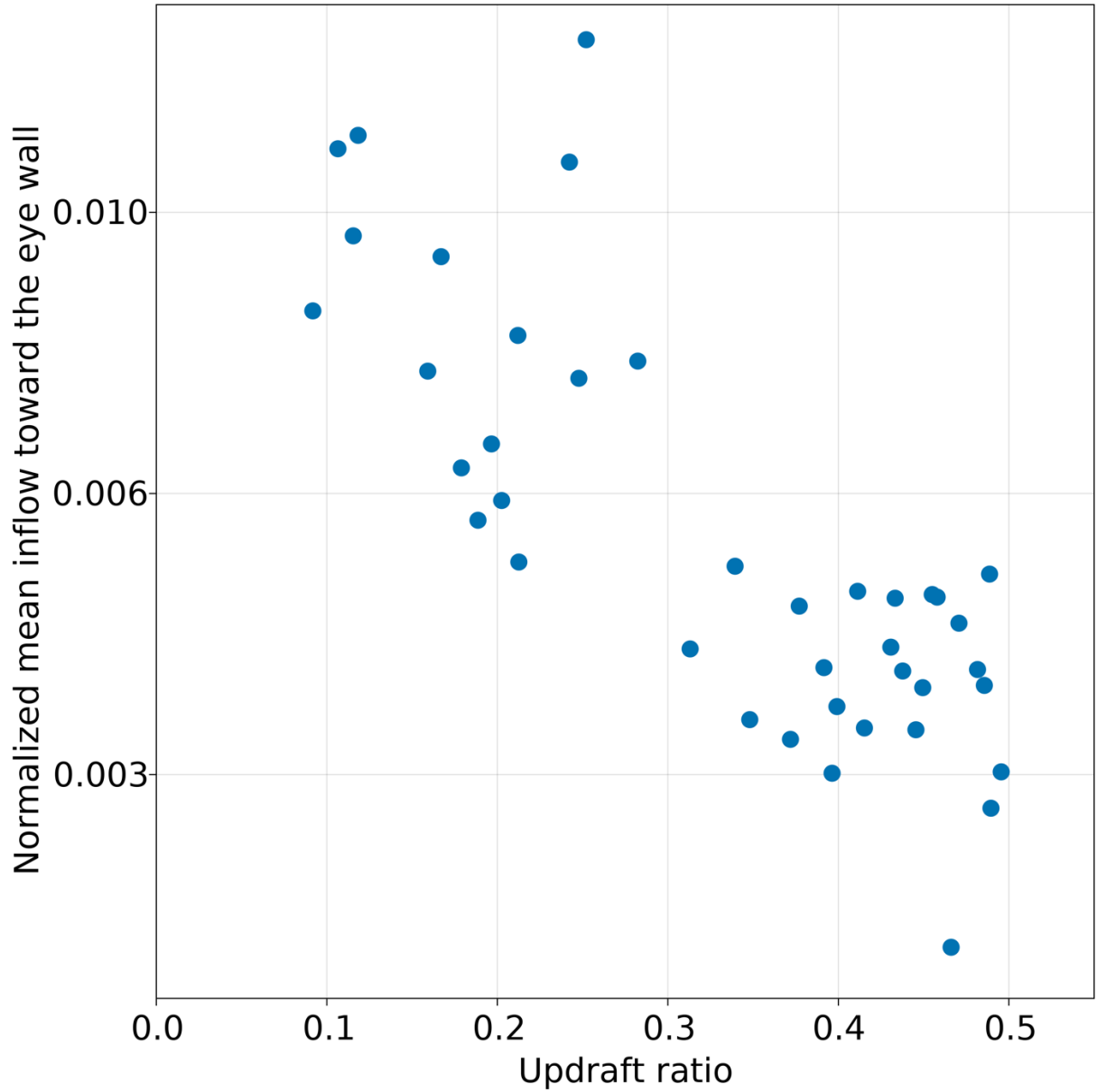

Figure 4. The absolute value of mean radial velocity from $= 0$ to $r_{max}$ and from surface to z = 6 km, normalized by the maximum tangential wind speed vs. the updraft ratio. The vertical axis has a logarithmic scale.